\documentclass[aps, twocolumn, superscriptaddress]{revtex4} 

\usepackage[switch]{lineno}
\usepackage{amsmath}
\usepackage{amssymb}
\usepackage{empheq}
\usepackage[parfill]{parskip}
\usepackage[active]{srcltx}
\usepackage{color}
\usepackage{array}
\usepackage{booktabs}
\usepackage{amsfonts}
\usepackage{dsfont}
\usepackage{graphicx}

\begin{document}

\setlength{\parindent}{0.5cm}

\title{Anti-phase synchronization in a population of swarmalators}

\author{Samali Ghosh}
\email{samalighosh816@gmail.com}
\affiliation{Physics and Applied Mathematics Unit, Indian Statistical Institute, 203 B. T. Road, Kolkata 700108, India}

\author{Gourab Kumar Sar}
\email{mr.gksar@gmail.com}
\affiliation{Physics and Applied Mathematics Unit, Indian Statistical Institute, 203 B. T. Road, Kolkata 700108, India}

\author{Soumen Majhi}
\email{soumen.majhi91@gmail.com}
\affiliation{Department of Ecology and Evolution, University of Chicago, Chicago, Illinois 60637, USA}

\author{Dibakar Ghosh}
\email{diba.ghosh@gmail.com}
\affiliation{Physics and Applied Mathematics Unit, Indian Statistical Institute, 203 B. T. Road, Kolkata 700108, India}

\begin{abstract}
\hspace{1 cm}  (Received XX MONTH XX; accepted XX MONTH XX; published XX MONTH XX) \\ 
Swarmalators are oscillatory systems endowed with a spatial component, whose spatial and phase dynamics affect each other. Such systems can demonstrate fascinating collective dynamics resembling many real-world processes. Through this work, we study a population of swarmalators where they are divided into different communities. The strengths of spatial attraction, repulsion as well as phase interaction differ from one group to another. Also, they vary from inter-community to intra-community. We encounter, as a result of variation in the phase coupling strength, different routes to achieve the static synchronization state by choosing several parameter combinations. We observe that when the inter-community phase coupling strength is sufficiently large, swarmalators settle in the static synchronization state. On the other hand, with a significant small phase coupling strength the state of anti-phase synchronization as well as chimera-like coexistence of sync and async are realized. Apart from rigorous numerical results, we have been successful to provide semi-analytical treatment for the existence and stability of global static sync and the anti-phase sync states.\\ 
\noindent\\
DOI: XXXXXXX
\end{abstract}

\maketitle
%%%%%%%%%%%%%%%%%%%%%%%%%%%%%%%%%%%%%
\section{Introduction}
Synchronization~\cite{pecora1990synchronization,boccaletti2002synchronization,arenas2008synchronization} refers to the phenomenon in which interacting dynamical systems adjust their rhythm in time, and the study of diverse aspects of this process has been at the forefront of research in nonlinear dynamics of networked systems. This is mainly due to its applicability in social and physical methods to biological and technological systems ~\cite{rosenblum2003synchronization,arenas2008synchronization}. From opinion formation~\cite{jusup2022social}, flashing of ﬁreﬂies~\cite{buck1988synchronous} to Josephson junction~\cite{wiesenfeld1996synchronization} or firing neurons~\cite{montbrio2015macroscopic}, synchronization takes place in a variety of natural and man-made systems.
%Researchers have predominantly concentrated on the incarnation and stability of variants of synchronous states emerging in coupled oscillatory systems. From single-layer and static topological frameworks~\cite{arenas2008synchronization} to multi-layer and temporal networks~\cite{ghosh2022synchronized}, this dynamics has, thus far, been investigated in many several different network configurations. 
 Alternative to the synchrony in the states of the systems, self-organization in space takes place in flocking birds, school of fish, swarming insects, a herd of sheep~\cite{buhl2006disorder,couzin2007collective,ballerini2008interaction,bialek2012statistical,leonard2012decision,garcimartin2015flow} and even in micro-organisms~\cite{zhang2010collective,levy2008stochastic,tian2021collective}. This phenomenon of spatial self-organization without explicit alteration in the internal states is referred to as the swarming~\cite{fetecau2011swarm,darnton2010dynamics,mogilner1999non}. On the contrary, as mentioned earlier, the internal state dynamics play the primary role in synchronization in which spatial movement does not necessarily participate. In the last few decades, these two fields, synchrony, and swarming, have been studied independently and parallel to each other. The study of mobile oscillators or moving agents brought these two fields into contact by considering the effect of oscillators' motion on the internal dynamics~\cite{frasca2008synchronization,majhi2019emergence}. In nature, \textcolor{black}{there are many cases of this behavior such as} vinegar eels~\cite{peshkov2022synchronized}, Japanese tree frogs~\cite{aihara2014spatio}, starfish embryos~\cite{tan2022odd} where the spatial and internal dynamics are dependent on each other. Recently, the combined effect of sync and swarming is expressed by a coupled system, namely swarmalators. The swarmalators~\cite{o2017oscillators,sar2022dynamics,yoon2022sync} are systems with concurrent existence of synchronization and swarming dynamics, particularly oscillatory systems having spatial and phase dynamics coupled. The study of such a fascinating interplay between the internal states and the positions in space was initiated by Tanaka et al.~\cite{tanaka2007general,iwasa2010hierarchical} who presented a swarm-oscillator model, followed by the recent swarmalator model proposed by O'Keeffe et al.~\cite{o2017oscillators}. The latter model built upon a space-dependent generalized version of globally coupled Kuramoto oscillators predicts few novel collective dynamics.%, namely static sync, static async, static phase wave, splintered phase wave, and  active phase wave.   

\par Significant efforts have been made since then in the last few years in order to understand the dynamics of swarmalators under different system-interaction setups. Phase similarity arising through both spatial attraction and spatial repulsion can result in ring phase wave states~\cite{o2018ring}. External periodic forcing affecting the phases leads to phase transitions from the states of the non-forced model through partial to full synchrony~\cite{lizarraga2020synchronization}. Besides the impact of thermal noise on the swarmalator system~\cite{hong2023swarmalators}, a number of other aspects, such as distributed coupling~\cite{o2022swarmalators}, time-delayed interactions~\cite{blum2022swarmalators}, finite-cutoff interaction distance~\cite{lee2021collective} are examined. The outcomes are a plethora of collective patterns, some observable in as disparate as Japanese tree frogs and electroporated Quincke rollers. Sar et al.~\cite{sar2022swarmalators} has come up with a swarmalator model subject to time-varying competitive phase interaction in which the competition between the attractive and repulsive interaction takes place depending on the sensing radius of the units.
%As a result, besides the active mixed phase wave, they observed a static $\pi$ state in which the system resides in a stationary two-cluster synchronized state with a phase difference of $\pi$ between the clusters.
Swarmalators on a ring subject to random pinning are investigated~\cite{sar2023pinning} and found to result in low-dimensional chaos, an abrupt transition to synchronous state, along with phase wave and split phase wave. Lately, Ceron et al.~\cite{ceron2023diverse} demonstrates that the edition of non-identical frequencies of the oscillators, local coupling, and chirality lead to new dynamics including beating clusters and lattices of vortices. Nevertheless, research in this fascinating world of swarmalators is still in its infancy and there are adequate scopes of further investigation leading to the possible revelation of new emerging collective dynamics due to the bidirectional reciprocity between the phase and the spatial dynamics.

\par One of the most pivotal characteristics of many real-world networked systems is that of community structures or clustering~\cite{girvan2002community,fortunato2010community} referring to the compartmental subdivisions of networked systems. This, precisely, corresponds to the organization of the units of the system in strongly intra-connected communities or groups while possessing weaker inter-group connections. From numerous social systems including collaboration networks to biological networks, such as metabolic networks, regulatory networks, and food webs, are naturally found to exhibit community structures~\cite{girvan2002community,fortunato2010community,newman2006modularity}. The problem of detection and characterization of these communities~\cite{fortunato2010community,malliaros2013clustering,clauset2004finding,newman2004finding} is one of the preeminent issues in the study of structural network theory. Through this article, we assume a community-structured framework of the underlying network and demonstrate the genesis of multiple variants of collective patterns in interacting communities of swarmalators. We, specifically, study a population of swarmalators where they are distributed in two communities. We analyze how the trade-off between the intra- and inter-community interactions affects the generic interplay between the phase and spatial dynamics of swarmalators. The phase interactions along with the spatial attraction and repulsion differ in each community. Under such a network setup, we encounter diverse routes to the static synchronization state as the inter-community phase coupling strength increases, for different choices of parameter values. Besides the states like active and static async or active phase wave, we detect anti-phase synchrony and the chimera state in the process towards the emergence of in-phase synchronization. We must here emphasize the remarkable fact that the anti-phase synchronization state arises in the sole presence of repulsive coupling even when the network size is considerably large, which we do not experience in the case of simple phase oscillator models without any spatial dynamics~\cite{tsimring2005repulsive}. We have also provided semi-analytical treatment concerning the stability analysis of the global static synchronization and the anti-phase synchronization state. We should mention here that, in a stereotypical community network structure, the strength of inter-community interaction is usually considered to be smaller than that of the intra-community interaction strength~\cite{girvan2002community}. In this work, however, we have not necessarily followed this convention. We have varied the inter-community phase coupling strength over a feasible range where a number of diverse collective states are observed.

%\par The rest of the paper is arranged as follows. In Sec.~\ref{section2}, we introduce and discuss the swarmalator model with an endowed community network structure. In Sec.~\ref{section3}, we define appropriate order parameters which are needed to study the emerging states. With the knowledge of the order parameters, we study different collective states of our model by varying the system parameters. We establish semi-analytical conditions for the realization of static sync and anti-phase states in Sec.~\ref{section 4}. The chimera state is demonstrated in detail in Sec.~\ref{chimerastate}. Finally, in Sec.~\ref{conclusion}, we summarize our findings and put forward concluding remarks.

\section{Proposed mathematical model}
\label{section2}
We consider $N$ number of swarmalators moving in a two-dimensional region. We randomly distribute them in $p$ groups. Let $C_i$ denote the set of indices of swarmalators belonging to the $i$-th group. Then trivially we have, $\sum_{i=1}^{p}|C_i| = N$, where $|C_i|$ denotes the cardinality of the set $C_i$ and $\cup_{i=1}^{p} C_i = \{1,2,\cdots,N\}$. Suppose, without loss of generality, that the $i$-th swarmalator belongs to the $n$-th group. Then we can write the governing equations as,
\begin{align}
    \Dot{\mathbf{x}}_i=\mathbf{v}_i &+ \sum_{m = 1}^{p}\dfrac{1}{|C_m|}\sum_{j \in C_m\setminus \{i \}}\Bigg[\dfrac{\mathbf{x}_j-\mathbf{x}_i}{|\mathbf{x}_j-\mathbf{x}_i|}\big(1 + \nonumber \\ &J_{n,m} \cos(\theta_j - \theta_i)\big)- \dfrac{\mathbf{x}_j-\mathbf{x}_i}{|\mathbf{x}_j-\mathbf{x}_i|^2}\Bigg],
    \label{eq1}
\end{align}
\begin{align}
    \Dot{\theta}_i=\omega_i + \sum_{m = 1}^{p}\dfrac{K_{n,m}}{|C_m|}\sum_{j \in C_m\setminus \{i \}} \dfrac{\sin(\theta_j-\theta_i)}{|\mathbf{x}_j-\mathbf{x}_i|},
    \label{eq2}
\end{align}
where $i=1,2,\cdots,N$. $\textbf{x}_i\equiv (x_i,y_i)$ is the spatial position in two-dimensional plane and $\theta_i$ is the internal phase of the $i$-th swarmalator. $\omega_i$ and $\mathbf{v}_i$ are the natural frequency, and self-propulsion velocity of the $i$-th swarmalator, respectively. The spatial attraction, repulsion as well as phase interaction functions are chosen the same as in Ref.~\onlinecite{o2017oscillators} where all the swarmalators belong to a single group, i.e., $p=1$. \textcolor{black}{The spatial attraction term ensures that the swarmalators remain close to each other without dispersing indefinitely, whereas spatial repulsion among them is necessary to avoid collision. They can be perceived as long-range attraction and short-range repulsion.} $J_{n,m}$ highlights how phases of those two swarmalators affect their spatial attraction. We assume $J_{n,m} > 0$ so that swarmalators which are in nearby phases attract each other spatially due to the presence of the term $\cos(\theta_j - \theta_i)$. Similarly, $K_{n,m}$ indicates the phase coupling strength between the two groups $C_n$ and $C_m$ (note that, here by group $C_n$, we mean the swarmalators belonging to the $n$-th group, without ambiguity). When $K_{n,m}>0$, swarmalators' phases are attractively coupled and the phase coupling is repulsive when $K_{n,m}<0$. For symmetry, $J_{n,m} = J_{m,n}$, and $K_{n,m}=K_{m,n}$. Then, for $p$ groups, the number of distinct parameters related to $J$ and $K$ are $(p^2+p)/2$ each. We work with $p=2$ groups in this article which leaves us with $J_{1,1}\equiv J_1$, $J_{2,2}\equiv J_2$, $J_{1,2}=J_{2,1}\equiv J_3$, $K_{1,1}\equiv K_1$, $K_{2,2}\equiv K_2$, and $K_{1,2}=K_{2,1}\equiv K_3$, say. Effectively we have these six parameters in hand which we vary to obtain different collective behaviors. Also note that, the model defined by Eqs.~\eqref{eq1}-\eqref{eq2} is a generalization of the model proposed by O'Keeffe et al.~\cite{o2017oscillators}. We work with swarmalators having identical natural frequencies and velocities, i.e., $\omega_i = \omega$ and $\mathbf{v}_i = \mathbf{v}$ for all $i$. By moving to a proper reference frame, we set $\omega=|\mathbf{v}| =0$.

\section{Results}
\label{section3}
First, we assume that the swarmalators are distributed in equal numbers in two populations (we remove this assumption later in Appendix~\ref{appendixa} to show that the results do not change if they are distributed unequally as long as $N$ is sufficiently large). For simplicity, let $C_i$ denote both the $i$-th population and the set of indices of swarmalators belonging to that population, whenever appropriate, for $i=1,2$. Then $J_1$ measures the extent to which the phases of swarmalators belonging to $C_1$ affect their spatial attraction and similarly $J_2$ for the group $C_2$. $J_3$ gauges the phase-dependent spatial attraction when swarmalators belong to different groups. $K_1$, and $K_2$ are the phase coupling strengths between swarmalators in $C_1$ and $C_2$, respectively, whereas $K_3$ is the strength of phase interaction when they are in different groups. The values of these control parameters decide the fate of the swarmalator system where we observe various emerging collective states by changing these values. Before moving forward to describe these states, first, we define some order parameters that are useful to measure several properties of the emerging states.
\subsection{Order parameters}
To measure the amount of synchrony in swarmalators' phases throughout the population, we define
\begin{equation} 
    r e^{i \psi} = \frac{1}{N} \sum_{j=1}^{N} e^{i\theta_j}.
    \label{r}
\end{equation}
Here $r$ lies between $0$ to $1$ by definition and gives an indication of the overall synchrony in swarmalators' phases. Phases are completely synchronized when $r=1$, or else asynchronous behavior is present. $\psi$ is the mean phase of the overall population. We measure the phase coherence among swarmalators belonging to the $p$-th group by
\begin{equation}
    r_p e^{i \psi_p} = \frac{1}{|C_p|} \sum_{j \in C_p} e^{i\theta_j},
    \label{rp}
\end{equation}
where $r_p$ again lies between $0$ to $1$ and $\psi_p$ is the average phase of $p$-th group. We also define
\begin{equation} \label{eq5}
    R e^{i \Psi} = \frac{1}{N} \sum_{j=1}^{N} e^{2 i\theta_j},
\end{equation}
which is useful to examine anti-phase synchrony where a phase difference of $\pi$ is observed among swarmalators' phases. In the anti-phase synchrony state, $R=1$ but $r\ne 1$. In some of the collective states (discussed later in Sec.~\ref{emergingstates}) we observe a correlation between swarmalators' phases $\theta_j$ and their spatial angle $\phi_j= \tan^{-1}(y_j/x_j)$. For this, we define the following order parameters,
\begin{equation}
    S_\pm e^{i \Psi \pm} = \frac{1}{N} \sum_{j=1}^{N} e^{i (\phi_j \pm \theta_j)},
    \label{s}
\end{equation}
which quantifies the correlation between phases and spatial angles. We take the maximum of $S_{\pm}$ and define $S=\max\{S_+,S_-\}$. A nonzero value of $S$ indicates the presence of a correlation between swarmalators' spatial angles and phases. In one of the collective states, swarmalators arrange themselves inside an annular-like structure and they rotate around this annulus. Their phases also keep changing from $0$ to $2\pi$. To distinguish this state from others, $\gamma$ is defined as
\begin{equation}
    \gamma = \frac{N_{rot}}{N},
    \label{gamma}
\end{equation}
where $N_{rot}$ is the number of swarmalators executing at least one full circle rotation in both spacial location and phase. $\gamma$ gives the fraction of such swarmalators and subsequently lies between $0$ to $1$. We find both stationary and non-stationary states in our model where swarmalators become static in position and phase in the stationary states but keep moving in the non-stationary ones. To separate these, we measure the mean velocity denoted by $V$, and is defined as
	\begin{equation}
	V = \Bigg \langle\frac{1}{N} \sum_{i=1}^{N} \sqrt{\dot{x}_i^2 + \dot{y}_i^2 + \dot{\theta}_i^2} \Bigg \rangle_t,
	\label{V}
	\end{equation}
where $\langle \cdots\rangle_t$ stands for the time average, which is taken after discarding the initial transients. With the knowledge of these order parameters, we proceed to study the emerging collective states of our model.

\begin{figure*}
    \centering
    \includegraphics[width=2\columnwidth]{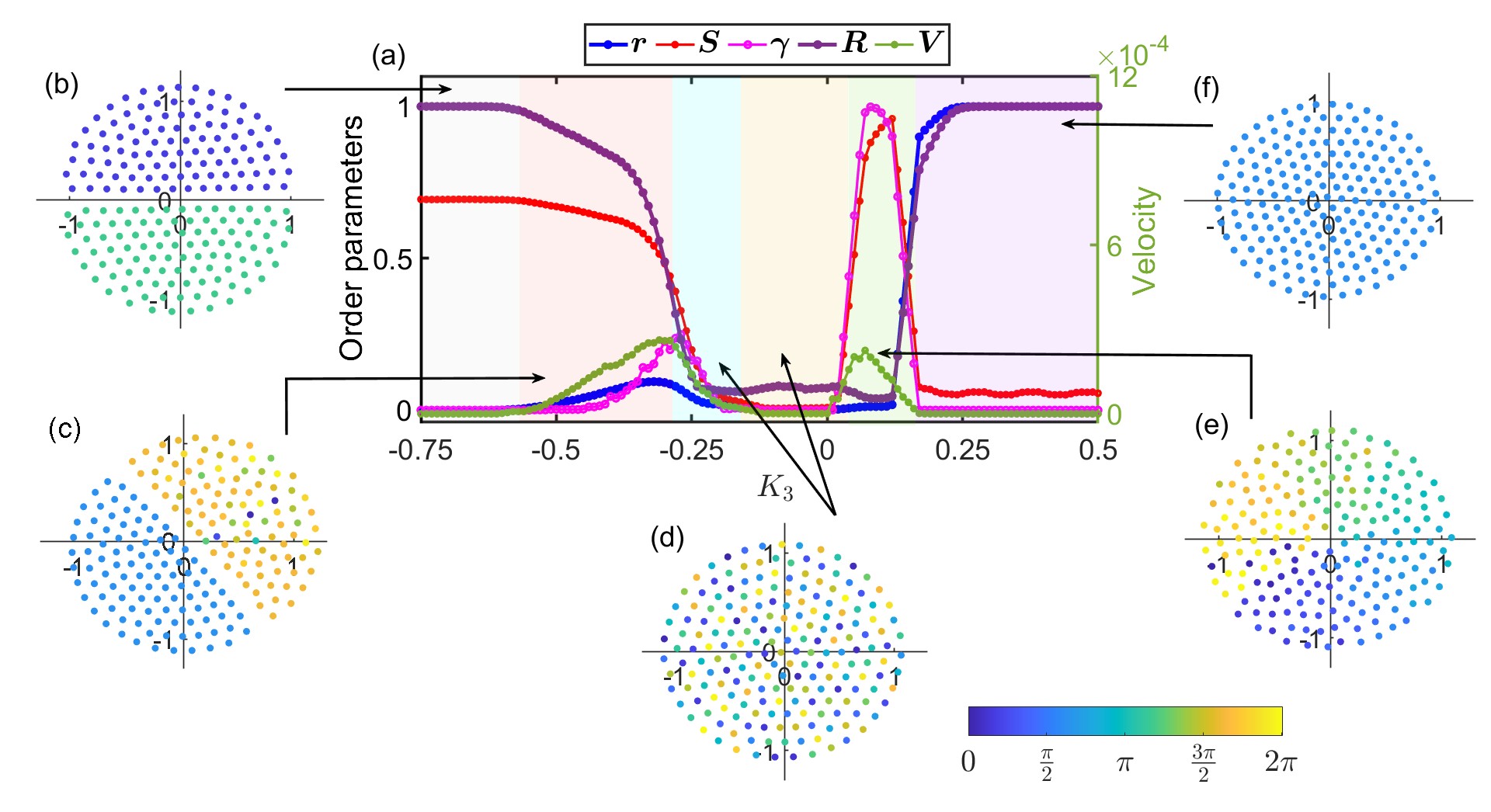}
    \caption{{\bf Order parameters along with the snapshots of the emerging states.} (a) Variation of different order parameters with $K_3$. (b) Anti-phase sync for $K_3=-0.6$, (c) chimera for $K_3=-0.4$, (d) active async for $K_3=-0.25$ \& static async for $K_3=0.0$, (e) active phase wave for $K_3=0.1$  and (f) static sync for $K_3=0.3$. Simulations are performed for $N=200$ swarmalators for $T=5000$ time units and step-size $dt=0.01$ using Heun's method. In all cases, swarmalators are initially placed inside the box $[-1,1] \times [-1,1]$ uniformly at random, while their phases are drawn randomly from $[0, 2\pi]$. We fix $J_1=J_2=J_3=0.1$ and $K_1=-0.1, K_2=-0.2$. Note that there is a long transient until the states are achieved. The order parameters are calculated with the last $10\%$ data after discarding the initial transients.} 
    \label{pt1}
\end{figure*}

\begin{table*}[ht]
		\caption{\label{table1}This table shows how the emerging states of the population of swarmalators are identified with the order parameters $r$, $R$, $S$, $\gamma$, and $V$. }
%		\begin{ruledtabular}
\centering
			\begin{tabular}{cccccc}
				
				$r$\hspace{0.5 cm} & $R$\hspace{0.5 cm} & $S$\hspace{0.5 cm} & $\gamma$\hspace{0.5 cm} & $V$\hspace{0.5 cm} & Emerging state\\ \hline
				$\approx 1$\hspace{0.5 cm} &$\approx 1$\hspace{0.5 cm} &$0<S\ll 1$\hspace{0.5 cm} &$\approx 0$\hspace{0.5 cm} &$\approx 0$\hspace{0.5 cm} & Static sync \\
    
                $\approx 0$\hspace{0.5 cm} &$0<R\ll1$\hspace{0.5 cm} &$\approx 0$\hspace{0.5 cm} &$\approx 0$\hspace{0.5 cm} &$\approx 0$\hspace{0.5 cm} & Static async\\
                
                $\approx 0$\hspace{0.5 cm} &$0<R\ll1$\hspace{0.5 cm} &$0<S\ll1$\hspace{0.5 cm} &$\ll 1$\hspace{0.5 cm} &$\ne 0$\hspace{0.5 cm} & Active async\\
                
                $\approx 0$\hspace{0.5 cm} &$0<R\ll1$\hspace{0.5 cm} &$0\ll S<1$\hspace{0.5 cm} &$0\ll\gamma <1$\hspace{0.5 cm} &$\ne 0$\hspace{0.5 cm} & Active phase wave\\

                $0<r\ll 1$\hspace{0.5 cm} &$0\ll R<1$\hspace{0.5 cm} &$0\ll S<1$\hspace{0.5 cm} &$\ne 0$\hspace{0.5 cm} &$\ne 0$\hspace{0.5 cm} & Chimera\\
                
			$\approx 0$\hspace{0.5 cm} &$\approx 1$\hspace{0.5 cm} &$0<S<R$\hspace{0.5 cm} &$\approx 0$\hspace{0.5 cm} &$\approx 0$\hspace{0.5 cm} & Anti-phase sync\\	
	\end{tabular}
	\end{table*}

\subsection{Emerging collective states}
\label{emergingstates}
We investigate the twin activities of synchronization and swarming in our model. For simplistic purpose, we take the values of $J_1$, $J_2$, and $J_3$ to be equal to $0.1$ and fix $K_1=-0.1$, $K_2=-0.2$. These choices of parameter values are arbitrary and solely made for a case study of our model. We relax this choice in the subsequent sections. However, the natural indication after performing numerical simulations is that $K_3$ is the most crucial parameter which determines the inter-group phase coupling. That is why we keep it as a free parameter and study our model's collective states while varying it. The model exhibits six long-term collective states: {\it anti-phase sync}, {\it chimera state}, {\it active async}, {\it static async}, {\it active phase wave}, and {\it static sync} when we vary $K_3$ inside an interval $[-0.75,0.5]$. Figure~\ref{pt1}(b)-(f) display the states by scatter plots in the $(x,y)$ plane where the swarmalators are represented by dots and they are colored according to their phases $\theta$. Figure~\ref{pt1}(a) reveals the variation of order parameters as a function of $K_3$. The order parameters $r, S,\gamma, R, V$ are plotted by blue, red, magenta, purple, and green-colored dotted lines, respectively. Table~\ref{table1} provides information regarding the values of the order parameters in these states. Next, we discuss these collective states and their structural properties in detail.

\par We start from the left end point of the interval. Here, $K_3 \ll K_1,K_2$. The population breaks into two disjoint clusters formed by the two groups of swarmalators. Both clusters are stationary in spatial position and phase. Swarmalators inside each cluster are completely synchronized. But, one cluster is synchronized at a common phase which is at $\pi$ difference from the common phase of the other cluster. We call this as {\it anti-phase sync} (see Fig.~\ref{pt1}(b)). Look at the white region of the parameter space of Fig.~\ref{pt1}(a), whereby the definition of $R$ in Eq.~\ref{eq5}, $R\approx 1$ in this state (purple curve). Since the overall population's phases are distributed in $\pi$ difference in two sub-populations and they are equal in size, we get $r \approx 0$ (blue curve). Being a static state, it also gives $\gamma \approx 0$ (magenta curve) and $V \approx 0$ (green curve). The other order parameter $S$ holds a nonzero value that is less than $R$ in this state (red curve). For a compact view of the order parameters, we refer the reader to Table~\ref{table1}. Section~\ref{antiphasestate} presents this state in a more detailed way. See Movie 1 of the Supplementary Material for the time evolution of this state.

\par When we gradually increase $K_3$, swarmalators in one community remain fully synchronized in phase but in the other community, asynchrony starts to appear. The cluster formation in the anti-phase sync state remains intact here (Fig.~\ref{pt1}(c)). However, asynchrony in one cluster brings some activity inside that cluster in the sense that swarmalators now move. This state is best visualized when studied in terms of $r_1$ and $r_2$. The synchronized cluster gives $r_1 = 1$, and the desynchronized one gives $r_2 <1$. This co-existence of synchronized and desynchronized swarmalator communities is reminiscent of the chimera state found in the study of coupled oscillators~\cite{abrams2008solvable,abrams2004chimera,majhi2019chimera} and we simply name this state as {\it chimera state}. All the five order parameters $r, S,\gamma, R, V$ show nonzero values here (pink region in Fig.~\ref{pt1}). See Table~\ref{table1} for more details. We also discuss this state in detail in Sec.~\ref{chimerastate}. Movie 2 of the Supplementary Material demonstrates the time evolution of the chimera state.

\par On further increment of $K_3$ from the chimera state, we encounter the swarmalators moving and arranging themselves within a circular disc and their phases are totally incoherent, i.e., $r \approx 0$. The activity never dies and they keep moving in the two-dimensional plane which gives $V \ne 0$. This state is named as {\it active async} as the swarmalators maintain movement in the $(x,y)$ plane, and their phases are desynchronized. Find Table~\ref{table1} for the description of order parameters in this state. Also, observe the cyan region in Fig.~\ref{pt1}. The activity dies keeping the disc structure with the incoherent phase nature when $K_3$ is increased beyond this state. This is the static async state. The only difference between this state and the active async state is that $V=0$ is in this state. Static async state prevails over the yellow region in Fig.~\ref{pt1}. Figure~\ref{pt1}(d) represents a snapshot at a particular time instant for both these states. See Movies 3 \& 4 of the Supplementary Material for the time evolution to this state.

\par Moving to the right with increasing $K_3$ from the static async state, we observe another collective state where the swarmalators arrange themselves inside an annular ring and oscillate to achieve regular cycles in both phase and space. This state was termed as {\it active phase wave} in previous studies \cite{o2017oscillators}. A snapshot of this motion is best illustrated in Fig.~\ref{pt1}(e). By our definition, $\gamma$ is nonzero in this state. Find the green region of Fig.~\ref{pt1} for the occurrence of this state (Movie 5 of the Supplementary Material best describes this state). Finally, to the extreme right of this parameter region where $K_3$ is sufficiently large and positive, phases of the swarmalators throughout the population get synchronized and they form a disc structure in the plane. This previously reported state is known as the {\it static sync} \cite{o2017oscillators}. The value of $r$ is the maximum here which is observed by the blue curve in the purple region. $R$ is also close to $1$ here by definition. Figure~\ref{pt1}(f) illustrates a snapshot of this state (also see Movie 6 of the Supplementary Material). 

\par
\begin{figure}[htp]
    \centering
    \includegraphics[width=\columnwidth]{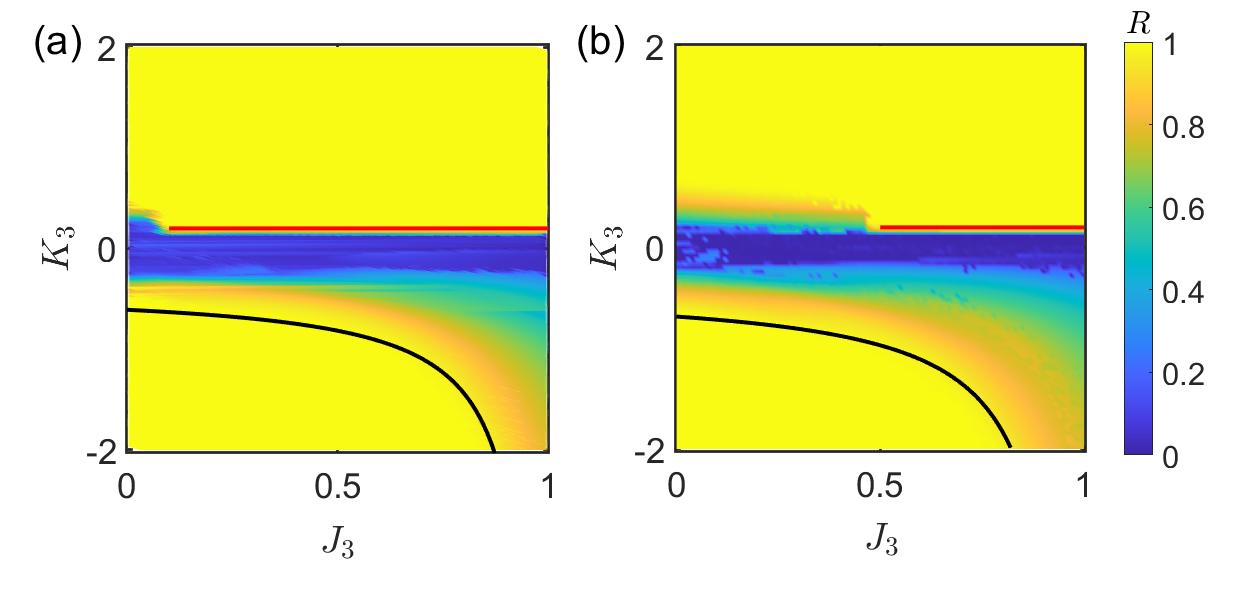}
    \caption{{\bf $J_3$-$K_3$ Parameter space for $J_1=J_2$.} (a) $J_1=J_2=0.1$. (b) $J_1=J_2=0.5$. The model is integrated with $N=200$ swarmalators using Heun's method with step-size $dt=0.01$ for $T=5000$ time units. Order parameter $R$ is calculated with the last $10\%$ data after discarding the transients. Colorbar stands for the value of $R$. Red and black curves are analytical predictions Eqs.~\eqref{cond1} and~\eqref{cond2}, respectively. Here $K_1=-0.1,K_2=-0.2$.}
    \label{ps}
\end{figure}

 Till now we have only varied $K_3$ and studied the emerging six collective states. Now, we simultaneously vary $J_3$ along with $K_3$ and observe the dynamical behaviors. The resulting parameter space is shown in Fig.~\ref{ps}. In this figure, the $J_3$-$K_3$ parameter plane is divided into $100\times 100$ mesh points, and at each point, we simulate our model for $T=5000$ time units. The value of order parameter $R$ is calculated over the last $10\%$ data and the mesh point is colored according to this value. We observe from Fig.~\ref{ps} that the emerging states are robust concerning variation in $J_3$. The top yellow region corresponds to the static sync state where $K_3$ is positive and $r, R \approx 1$. The yellow region towards the bottom corresponds to the anti-phase sync state where $R \approx 1$, but $r \approx 0$ (not shown here). The red and black curves are the analytical predictions for achieving the static sync and anti-phase sync state, respectively. Find Sec.~\ref{staticsyncstate} and~\ref{antiphasestate} for the derivation of these curves. So far we have always considered $J_1=J_2$. Appendix~\ref{appendixb} demonstrates the picture when we work with $J_1 \ne J_2$. The emerging states remain the same which can be seen from Fig.~\ref{j1-ne-j2}.

\section{Emerging collective states from identical communities: Dynamical routes}
We know from Ref.~\cite{o2017oscillators} that with a single community structure, our model exhibits five long-term states depending on the parameter values. These states are static sync, static async, static phase wave, splinter phase wave, and active phase wave of which the last two are non-stationary states. Here we assume that both communities are in the same state which belongs to one of these five states. This means the communities are identical with $J_1=J_2$ and $K_1=K_2$. Furthermore, we fix $J_3 = 0.1$ and analyze the routes from static anti-phase sync to static sync by varying the parameter $K_3$ over a range to perceive the collective states.

\begin{figure*}
    \centering
    \includegraphics[width=2\columnwidth]{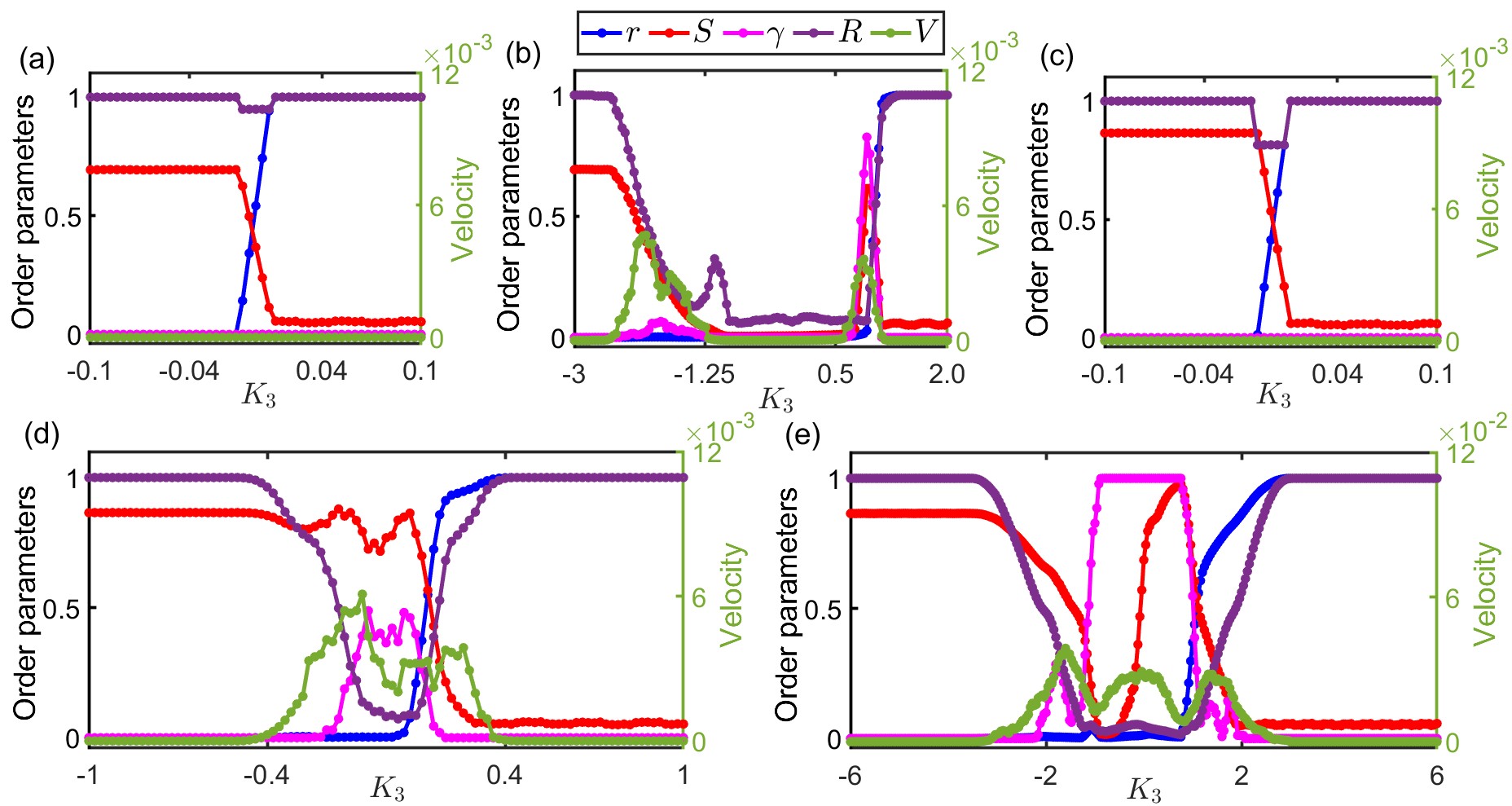}
    \caption{{\bf Behavior of the order parameters for identical swarmalator communities.} Order parameters as a function of $K_3$ where initially both the communities are in (a) static sync ($J_1=J_2=0.1$, $K_1=K_2=1.0$), (b) static async ($J_1=J_2=0.1$, $K_1=K_2=-1.0$), (c) static phase wave ($J_1=J_2=1.0$, $K_1=K_2=0.0$), (d) splintered phase wave ($J_1=J_2=1.0$, $K_1=K_2=-0.1$) and (e) active phase wave ($J_1=J_2=1.0$, $K_1=K_2=-0.75$). Simulation parameters $(dt,T,N)=(0.01,5000,200)$. The order parameters are calculated with the last $10\%$ data. Here, we fix $J_3=0.1$.} 
    \label{op_ppt}
\end{figure*}

\subsection{Static sync}
We start from the static sync state for both the communities ($J_1=J_2=0.1$, $K_1=K_2=1.0$) and change $K_3$. Firstly, in the negative $K_3$ region, we notice the population forms two clusters that are static in both phase and position. They are separated by a phase difference of $\pi$ from each other, which is the anti-phase sync state. An increment of $K_3$ shows the persistence of cluster structure but with a lower phase difference (see Movie 7 of the Supplementary Material). This cluster synchronization state (which is not the anti-phase sync state) exists over a small interval of $K_3$ before finally yielding the static sync state. With increasing $K_3$, we find \\\\
anti-phase sync $\rightarrow$ cluster sync $\rightarrow$ static sync.\\\\ See Fig.~\ref{op_ppt}(a) for the behavior of order parameters here.

\subsection{Static async}
Here, we take the two groups initially at the static async state by choosing the parameter values as $J_1=J_2=0.1$ and $K_1=K_2=-1.0$ and vary $K_3$. Starting from a relatively lower value of $K_3$ in comparison to $K_1$ and $K_2$, we notice the presence of an anti-phase sync state where two synchronized, stable clusters maintain a phase difference of $\pi$. When we increase $K_3$ in very small magnitude, we observe the chimera state where one group of swarmalators are fully phase coherent and in the other group they are out of synchrony. We encounter the active async state as we further increase $K_3$. After this, activity dies and the swarmalators arrange themselves in a static async by adjusting their spatial position with further increments of $K_3$. From this, we spot the emergence of an active phase wave state by increasing $K_3$. As $K_3$ is further increased, the whole community accomplishes themselves in static sync finally. Figure~\ref{op_ppt}(b) portrays the phase transitions in this case. The route is\\\\ 
anti-phase sync $\rightarrow$ chimera $\rightarrow$ active async $\rightarrow$ static async $\rightarrow$ active phase wave $\rightarrow$ static sync.\\\\ Compared to the earlier case where both the communities were in a static sync state, here we observe that the intermediate dynamics are relatively richer when the communities are in static async.

\subsection{Static phase wave}
Here both communities are in a static phase wave state. Primarily, here we deal with a phase-dependent aggregation model as $J_1=J_2=1.0$ and $K_1=K_2=0$. So, intra-community phase coupling is absent here. Phase interaction only takes place through the inter-community structure via $K_3$. We can trace the anti-phase sync in the negative $K_3$ region as before. The swarmalators follow a path from anti-phase sync to static sync through a static phase wave state which is deformed in nature, i.e., they are distributed in a non-uniform pattern in the $2$D plane (in the existing static phase wave state, they are distributed uniformly in an annular ring). We can trace this deformed state very close to the $K_3 = 0$ region. Movie 8 of the Supplementary Material best describes this state. Swarmalators arrange themselves into static sync for $K_3>0$. Here, the route can be noted down as:\\\\ 
anti-phase sync $\rightarrow$ deformed static phase wave $\rightarrow$ static sync.\\\\ The order parameters can be found in Fig.~\ref{op_ppt}(c).

\subsection{Splintered phase wave}
So far we deal with the scenario where both the communities are in static states initially. Here we start with two identical non-stationary states namely splinter phase wave. We keep the parameter values $J_1=J_2=1.0$ and $K_1=K_2=-0.1$ and vary $K_3$. We observe anti-phase sync where two static, synchronized clusters persist with a phase difference $\pi$ for a relatively smaller value of $K_3$ compared to $K_1$ and $K_2$. Increasing the value of $K_3$, we mark the splintered phase wave state $-0.4<K_3<0.22$. Here the swarmalators split into two clusters where the mean phases of the clusters differ from each other by approximately $\pi$ (Movie 9 of the Supplementary Material). Moving to the right with an increasing value of $K_3$, we notice some of the swarmalators start to execute a full cycle rotation spatially but their phases do not change from $0$ to $2\pi$ as in the active phase wave state. This peculiar state can be deciphered as the simultaneous coexistence of splintered phase wave and active phase wave states (see Movie 10 of the Supplementary Material for an illustration of the state). We observe the mixed activity of splintered and active phase wave states when $-0.22<K_3<0.22$. Further increasing $K_3$, the swarmalators are again divided into two clusters but this time they maintain a difference of mean phases around $0$ (see the time evolution of this state in Movie 11 of the supplementary Material). This state exists for $0.22<K_3<0.4$. Finally, the whole population reaches static synchrony after a certain value of $K_3$ ($\approx 0.4$). The overall route is depicted as:\\\\ anti-phase sync $\rightarrow$ splintered phase wave (mean phase difference close to $\pi$) $\rightarrow$ mixed (splintered and active phase waves) $\rightarrow$ splintered phase wave (mean phase difference close to $0$) $\rightarrow$ static sync.

\subsection{Active phase wave}
Here the story starts with two identical active phase wave states. The parameter values are $J_1=J_2=1.0$ and $K_1=K_2=-0.75$. To analyze the route from anti-phase synchrony to static synchrony, we vary $K_3$ over a broad range. In this case, the anti-phase sync state is found for a relatively larger negative value of $K_3$ compared to the previous cases ($K_3 < -2.12$). Increasing the value of $K_3$, swarmalators start to segregate into two clusters and we observe activity emerging in the system. Some of the swarmalators undergo a full circle rotation in space and phase and consequently, $\gamma$ exhibits a small non-zero value around $-2.16<K_3<-1.28$. Movie 12 of the Supplementary Material demonstrates the state best. On further increment of $K_3$, we notice their oscillations increase in amplitude until all of them start to execute regular cycles in both phase and spatial angle, i.e., the swarmalators settle in the active phase wave state. The value of $\gamma$ is close to $1$ and $S$ is very small. With a further increment of $K_3$, their activity begins to diminish gradually and they are again separated into two clusters. The phase difference also decreases between the two clusters and $\gamma$ is very small compared to $1$ near $1.12<K_3<2.0$ (see Movie 13 of the Supplementary Material). Ultimately we find static sync for $K_3>2.0$. The route in this case becomes\\\\ anti-phase sync $\rightarrow$ mixed (splintered and active phase waves) $\rightarrow$ active phase wave $\rightarrow$ mixed (splintered and active phase waves) $\rightarrow$ static sync.

\section{Analytical findings}
\label{section4}
In the previous section, we explored the dynamic states of our model with various parameter values. The most striking result that we encountered is the occurrence of anti-phase sync with a reasonably small value of $K_3$ and on the other hand, a sufficiently large and positive value of $K_3$ results in the whole population in the static sync state. These two static states exist at the opposite extremes of $K_3$ values. Now, we try to establish the criteria for achieving these states.

\subsection{Static sync state} 
\label{staticsyncstate}
\begin{figure}[hpt]
    \centering
    \includegraphics[width=\columnwidth]{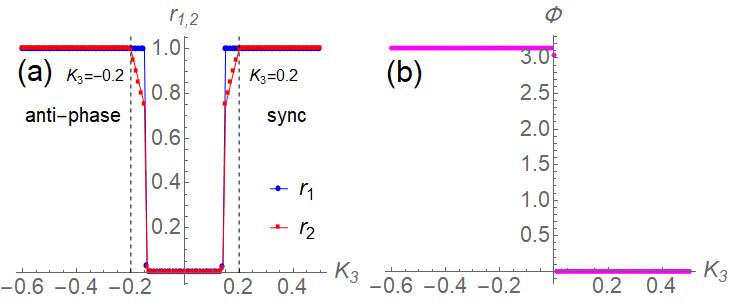}
    \caption{\textcolor{black}{{\bf Variation of order parameters $r_1$, $r_2$ and phase difference  $\Phi$ as functions of $K_3$.} We integrate Eqs.~\eqref{o1}-\eqref{o3} starting from initial conditions $(r_1(0),r_2(0),\Phi(0))=(0.9,0.9,\pi-0.1)$ for $T=5000$ time units. Then they are time averaged over the last $10\%$ data and plotted as functions of $K_3$. We fix $K_1=-0.1, K_2=-0.2$ like in Fig.~\ref{pt1}. (a) The variations of $r_1$ (blue) and $r_2$ (red) are plotted. (b) Represents the change of $\Phi$ (magenta) with varying $K_3$.}}
    \label{rev1}
\end{figure}

Before going into the study of the static sync state, we first analyze the phase dynamics of our model where spatial positions do not affect the phases. In that case, the phase equation becomes
\begin{align}
    \Dot{\theta}_i=\omega_i + \sum_{m = 1}^{2}\dfrac{K_{n,m}}{|C_m|}\sum_{j \in C_m\setminus \{i \}} \sin(\theta_j-\theta_i),
    \label{kuramoto}
\end{align}
where $i \in C_n$. We move to the continuum limit where $|C_p| \rightarrow \infty$, $p=1,2$. Considering the probability density function $\rho_n(\theta,t)$ of oscillators belonging to the $n$-th group, the Fokker-Planck equation can be written as
\begin{equation}
    \dfrac{\partial \rho_n}{\partial t} + \dfrac{\partial}{\partial \theta} (\rho_n v_n) = 0,
    \label{fp}
\end{equation}
where the velocity $v_n(\theta^n,t)$ is given by
\begin{equation}
    v_n(\theta^n,t) = \omega + \sum_{m=1}^{2} K_{n,m} \int \sin(\theta^m - \theta^n) \rho_m(\theta^m,t) d\theta^m.
    \label{velocity}
\end{equation}
We define the complex order parameter
\begin{equation}
    z_n(t) = \sum_{m=1}^{2} K_{n,m} \int e^{i\theta^m} \rho_m(\theta^m,t) d\theta^m.
    \label{eq12}
\end{equation}
Using this, Eq.~\eqref{velocity} is re-written as
\begin{equation}
    v_n(\theta^n,t) = \omega + \frac{1}{2i} (z_n e^{-i\theta^n} - z_n^* e^{i\theta^n}),
\end{equation}
where $*$ denotes complex conjugate. Following Ott-Antonsen ansatz \cite{ott2008low}, we choose a special class of density functions that has an invariant manifold of Poisson kernels,
\begin{equation}
    \rho_n(\theta^n,t) = \frac{1}{2\pi}\left\{ 1 + \left[ \sum_{k=1}^{\infty}[a_n(t)e^{i\theta^n}]^k + c.c.\right]\right\}.
    \label{eq14}
\end{equation}
where the unknown function $a_n(t)$ must be found self-consistently. Inserting this form of $\rho_n$ given by Eq.~\eqref{eq14} into Eq.~\eqref{fp}, we find that $\rho_n$ satisfies the Fokker-Planck equation for all harmonics $k$ if $a_n$ satisfies
\begin{equation}
    \dot{a}_n + i \omega a_n + \frac{1}{2}\left[a_n^2 z_n - z_n^*\right] = 0.
\end{equation}
Further inserting Eq.~\eqref{eq14} into Eq.~\eqref{eq12} and after performing the integration, the complex order parameter $z_n$ is expressed in terms of $a_n$ as
\begin{equation}
    z_n(t) = \sum_{m=1}^{2} K_{n,m} a_m^*(t).
\end{equation}
Then the amplitude equation for $a_1$ becomes
\begin{align}
    \Dot{a}_1 = &- i \omega a_1 - \frac{1}{2} (K_{1,1}a_1^* + K_{1,2} a_2^*) \nonumber \\&+\frac{1}{2} (K_{1,1}a_1 + K_{1,2} a_2).
    \label{eq17}
\end{align}
Similarly, we find the equation for $\Dot{a}_2$ by interchanging $1$'s and $2$'s in Eq.~\eqref{eq17}. We move to the polar coordinates to rewrite the amplitude equations by defining $a_n = r_n e^{-i \phi_n}$, $n=1,2$. We further define $\Phi = \phi_1 - \phi_2$. Substituting these into the amplitude equations and after simplifying, we get
\begin{align}
    \Dot{r_1} &= \frac{1-r_1^2}{2} (K_1 r_1+K_3 r_2 \cos \Phi ),\label{o1}\\
    \Dot{r_2} &= \frac{1-r_2^2}{2} (K_2 r_2+K_3 r_1 \cos \Phi ),\label{o2}\\
    \Dot{\Phi} &= -K_3  \left(\frac{r_1^2+r_2^2+2 r_1^2 r_2^2}{2 r_1 r_2}\right) \sin \Phi \label{o3}
\end{align}
 (note that, $K_{1,1} \equiv K_1$, $K_{2,2} \equiv K_2$, and $K_{1,2} = K_{2,1} \equiv K_3$). {\color{black} We integrate Eqs.~\eqref{o1}-\eqref{o3} with initial conditions $(r_1(0),r_2(0),\Phi(0))=(0.9,0.9,\pi-0.1)$ and demonstrate their variation as functions of $K_3$ in Fig.~\ref{rev1} where $K_1$ and $K_2$ are fixed to $-0.1$ and $-0.2$, respectively. When $K_3<-0.2$, $r_1=r_2=1$ and $\Phi=\pi$ which represents the anti-phase sync state (we study this state in detail in the next section). On the opposite side, for $K_3>0.2$, we get $r_1=r_2=1$ and $\Phi=0$ which stand for the sync state. In the middle region $-0.2<K_3<0.2$, chimera-like states appear.}

 \begin{figure*}
    \centering
    \includegraphics[width=2\columnwidth]{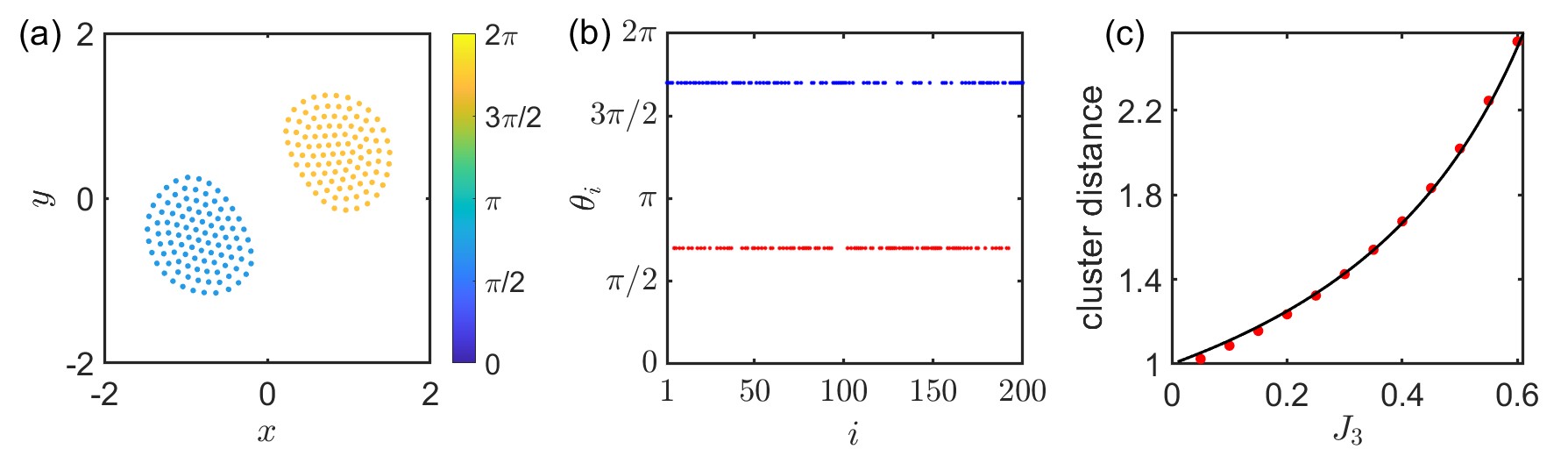}
    \caption{{\bf Anti-phase sync state.} The entire swarmalator population forms two disjoint clusters in space where the clusters belong to the two communities. Simulation parameters: $J_1=J_2=0.5$. $K_1=-0.1,K_2=-0.2,K_3=-1.5$. $(dt,T,N)=(0.01,5000,200)$. $J_3=0.5$ in (a) and (b). (a) Snapshot at $T=5000$ time units showing the spatial structure of the swarmalators in the anti-sync state where they are colored according to their phases. (b) The phases of the swarmalators are plotted against their respective indices at $T=5000$ time units where red and blue dots correspond to swarmalators belonging to the first and second communities, respectively. The phase difference is $\pi$ between the communities. (c) The distance between the center of masses of the two clusters is plotted as a function of $J_3$. Red dots are simulation results and the black curve indicates the analytical prediction, Eq.~\eqref{dist}.}
    \label{antiphase}
\end{figure*}
 \par In the global sync state, the phases of the swarmalators throughout the entire population become identical, which yields $r_1=r_2=1$ and $\Phi=0$. This is the trivial solution to Eqs.~\eqref{o1}-\eqref{o3} and the Jacobian matrix at this steady state gives eigenvalues $-2K_3, -K_1-K_3,-K_2-K_3$. From this, the sync state stability condition is achieved as
\begin{equation}
    K_3 > \max \{0,-K_1,-K_2\}.
    \label{cond1}
\end{equation}
When we consider the phase dynamics of swarmalators, Eq.~\eqref{eq2}, the spatial effect is to be dealt with. However, from numerical simulations, we observe that when $J_1=J_2$ and $J_3>J_1$, the stability condition Eq.~\eqref{cond1} holds for achieving the static sync state. This is demonstrated by the red lines in Fig.~\ref{ps}. For non-identical values of $J_1$ and $J_2$, the spatial distributions of swarmalators in the two communities do not remain the same. The spatial positions having an impact on the phase dynamics, in turn, affect the critical $K_3$ in Eq.~\eqref{cond1}. The small deviation from the condition given by Eq.~\eqref{cond1} (plotted by the red curve) is observed in Fig.~\ref{j1-ne-j2} (Appendix~\ref{appendixb}) where we present our results with $J_1=0.1$ and $J_2=0.5$.

\subsection{Anti-phase sync state}
\label{antiphasestate}
In the anti-phase sync state, the two groups get separated from each other in the phase component. Their phases are fully synchronized within each group, but there is a phase difference of $\pi$ between these two groups (see Fig.~\ref{antiphase}(b)). When $J_3$ is absent, i.e., $J_3=0$, these two groups arrange themselves spatially in the form of a disc where these discs overlap. The radius of these discs depends on the choices of $J_1$ and $J_2$. But when the value of $J_3$ is nonzero, swarmalators belonging to different groups start to reduce the attraction between them (since the strength of attraction between these two groups is $1-J_3$ as phase difference is exactly $\pi$). As a result, these two groups form disjoint clusters in the plane. See Fig.~\ref{antiphase}(a). 

In the anti-phase sync state, $r_1=r_2=1$ and $\Phi = \pm \pi$. These also satisfy Eqs.~\eqref{o1}-\eqref{o3} in the steady state. Linearizing these equations around this steady state and calculating the Jacobian matrix, yields the eigenvalues $2 K_3$, $K_3-K_1$, and $K_3-K_2$. This gives the stability condition of the anti-phase sync state as
\begin{equation}
    K_3 < \min \{0,K_1,K_2\}.
    \label{cond}
\end{equation}
We use Eq.~\eqref{cond} to find the stability condition of the anti-phase sync state found in our systems defined by Eqs~\eqref{eq1}-\eqref{eq2}. Since in our model, the phase dynamics of the swarmalators are influenced by the spatial dynamics, we first study this effect in the anti-phase sync state. From simulation results, we find that in the anti-phase sync state with nonzero $J_3$, swarmalators form disjoint clusters in a two-dimensional plane. Swarmalators belonging to group $C_1$ make a cluster among them where their phases are synchronized and the other cluster is formed by the swarmalators similarly belonging to $C_2$. This can be considered as a two-particle system where swarmalators belonging to the same group are represented by their center of positions and synchronized phase \cite{sar2022swarmalators}. Let $\mathbf{x}_{C_1}$ and $\mathbf{x}_{C_2}$ be the center of positions of $C_1$ and $C_2$ and $\theta_{C_1}$, $\theta_{C_2}$ be their synchronized phase angles, respectively. Then from Eq.~\eqref{eq1}, we can write
\begin{align}
    0=\Bigg[\dfrac{\mathbf{x}_{C_2}-\mathbf{x}_{C_1}}{|\mathbf{x}_{C_2}-\mathbf{x}_{C_1}|}\big(1 + J_3 \cos(\theta_{C_2} - \theta_{C_1})\big)- \dfrac{\mathbf{x}_{C_2}-\mathbf{x}_{C_1}}{|\mathbf{x}_{C_2}-\mathbf{x}_{C_1}|^2}\Bigg].
    \label{cps}
\end{align}
This gives us the distance between the center of positions of $C_1$ and $C_2$ as 
\begin{equation}
    |\mathbf{x}_{C_2}-\mathbf{x}_{C_1}| = \frac{1}{1-J_3},
    \label{dist}
\end{equation}
since $|\mathbf{x}_{C_2}-\mathbf{x}_{C_1}| \ne 0$ and $\theta_{C_2} - \theta_{C_1} = \pm \pi$. This is plotted in the black line in Fig.~\ref{antiphase}(c) where the red dots are simulation results. When the swarmalators form separate groups in spatial positions, their effective phase coupling strength changes since it depends on the distance between the swarmalators. This is why Eq.~\eqref{cond} does not stand valid for our model. To find the stability condition of the anti-phase sync state, we need to investigate the effect of spatial position carefully. The average distance $R_1$ between two particles in $C_1$ can be considered as the half of its diameter (maximum distance between particles in $C_1$) which is a function of $J_1$,$J_2$, and $J_3$, i.e., $R_1(J_1, J_2, J_3)$. Similarly for $C_2$ it is $R_2(J_1,J_2,J_3)$. On the other hand, the average distance of the particle throughout the whole population then becomes $R_1+R_2+1/(1-J_3) = R_3$, say. The effective ratio of $K_3$ to $K_1$ can be written down as $R_1 K_3/R_3 K_1$ and that of $K_3$ to $K_2$ is $R_2 K_3/R_3 K_1$. Then from Eq.~\eqref{cond}, we write down the stability condition of the anti-phase sync state as
\begin{equation}
    K_3 < \min \{0,\frac{R_3 K_1}{R_1},\frac{R_3 K_2}{R_2}\}.
    \label{cond2}
\end{equation}
Due to the complexity of the model, we are unable to find explicit expressions for $R_1$ and $R_2$. But from numerical simulations, we observe that these quantities depend majorly on the values of $J_1$ and $J_2$ for respective groups and not on $J_3$. We can approximately write $R_1 \approx R_1(J_1)$ and $R_2 \approx R_2(J_2)$. To verify our results, we take $J_1=J_2=0.1$ and $K_1=-0.1, K_2=-0.2$. Numerical simulations suggest $R_1 \approx 0.98 \approx R_2$. The curve defined by Eq.~\eqref{cond2} is drawn with these values and is plotted in black in Fig.~\ref{ps}(a). With $J_1=J_2=0.5$ and same $K_1$ and $K_2$ we find $R_1 \approx 0.7 \approx R_2$. The separatrix curve is again calculated and plotted in black in Fig.~\ref{ps}(b). Both curves match very well with our numerical results. We also verify our findings with unequal $J_1$ and $J_2$ in Appendix~\ref{appendixb}.

\section{Chimera state}
\label{chimerastate}
\begin{figure}[hpt]
    \includegraphics[width=\columnwidth]{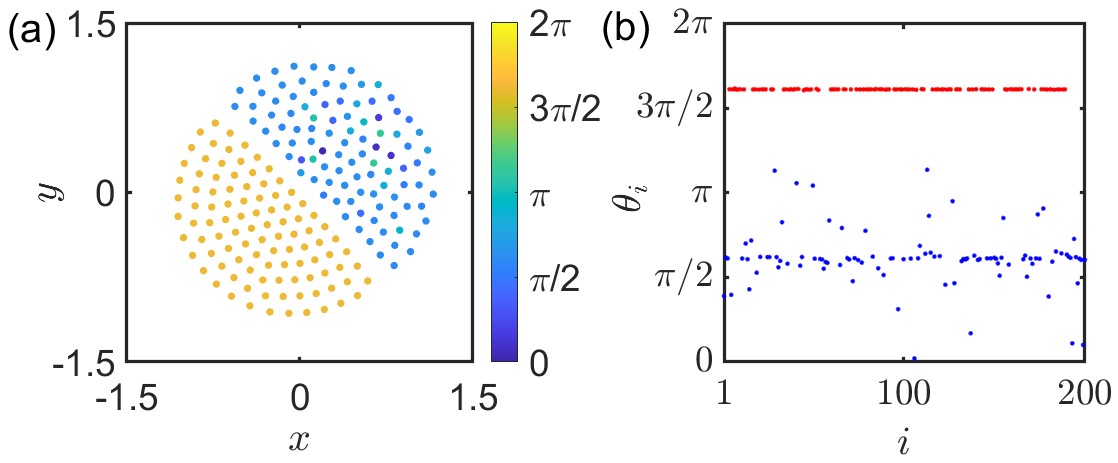}
    \caption{{\bf Chimera state.} One of the communities is completely phase synchronized and the presence of asynchrony is found in the other community. Simulation parameters: $J_1=J_2=J_3=0.1$. $K_1=-0.1$, $K_2=-0.2$, $K_3=-0.4$, $(dt,T,N) = (0.01,5000,200)$. (a) Snapshot at $T=5000$ time units demonstrating the chimera state. (b) Snapshots of the swarmalators' phases are plotted against their indices. The red and blue dots refer to the first and second communities, respectively. Swarmalators are synchronized in the first community (red dots) but desynchronized in the second one (blue dots).}
    \label{chimera1}
\end{figure}

The co-existence of coherence and incoherence is known as chimera state~\cite{abrams2008solvable,abrams2004chimera,majhi2019chimera}. We find that for certain parameter values, there is complete synchrony among one group of swarmalators but the other group is desynchronized. This means one of $r_1$ and $r_2$ is $1$ and the other one is strictly less than $1$. We display one such occurrence of chimera state in Fig.~\ref{chimera1}. Snapshot at $t=300$ time units with $N=200$ swarmalators is shown in Fig.~\ref{chimera1}(a) where the two groups arrange themselves in the $x$-$y$ plane in the shape of non-overlapping half discs. Here $r_1 = 1$ and $r_2 < 1$. This is evident when we look at the phases of the swarmalators in Fig.~\ref{chimera1}(b). We plot the phases of the swarmalators against their indices where red and blue dots stand for groups one and two, respectively.

\par
\begin{figure}[htp]
    \centering
    \includegraphics[width=\columnwidth]{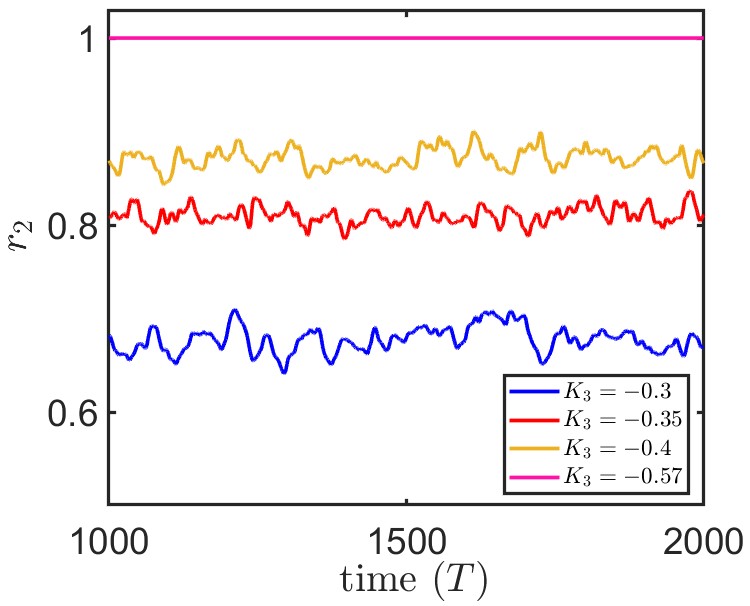}
    \caption{{\bf Breathing chimera}. We delineate the breathing nature of the chimera state. Parameter values used: $J_1=J_2=J_3=0.1$. $K_1=-0.1,K_2=-0.2$. Simulation parameters $(dt,T,N)=(0.01,2000,200)$. The order parameter $r_1$ measuring the phase coherence among the swarmalators in the first community acquires the value $1$. But $r_2$ is less than $1$ due to the presence of asynchrony in the second community. The temporal evolution of $r_2$ is plotted for several $K_3$ values, $K_3=-0.3$ (blue), $-0.35$ (red), $-0.4$ (yellow), and $-0.57$ (magenta). We observe oscillatory behavior of $r_2$ which reveals the breathing nature of the chimera state. The magnitude of $r_2$ increases and the oscillation decays with decreasing $K_3$ until it reaches the maximum value $1$ where the oscillation completely dies.}
    \label{chimera3}
\end{figure}

 We further study the nature of the chimera state. For a case study, we fix $J_1=J_2=J_3=0.1$ and set $K_1$ and $K_2$ to $-0.1$ and $-0.2$, respectively. By careful investigation, we find that a chimera state exists for these parameter values when $-0.56<K_3<-0.28$. In the chimera state, $r_1$ stays fixed to $1$ but $r_2$ is always less than $1$. Moreover, we observe oscillation in $r_2$, which means it varies with time. So, the chimera we report in this work is {\it breathing chimera}. We establish this by drawing Fig.~\ref{chimera3} where $r_2$ is plotted as a function of time for various values of $K_3$. It is to be noted that, with decreasing $K_3$ the magnitude of $r_2$ keeps increasing. Eventually around $K_3 \approx -0.57$, $r_2$ goes to $1$, which is the anti-phase sync state.

\section{Conclusion}
\label{conclusion}
The phase-dependent spatial aggregation and position-dependent phase synchronization are at the core of swarmalator dynamics. Swarmalators endowed with spatial and phase interactions are competent to exhibit complex collective behaviors. These states can be found in real-world systems like Japanese tree frogs~\cite{aihara2014spatio}, magnetic domain walls~\cite{hrabec2018velocity}, Janus matchsticks~\cite{chaudhary2014reconfigurable}, robotic swarms~\cite{schilcher2021swarmalators,barcis2020sandsbots} etc. To this end, studies are being carried out on swarmalator models by defining suitable interaction functions, network structures, coupling schemes, etc. (We refer the reader to~\cite{sar2022dynamics} for a recent review on swarmalator systems.)

\par In this article, we have studied a population of swarmalators where they are distributed in two communities. The intra and inter-community coupling strengths have been carefully varied to observe different emerging states. Two of them, viz., the anti-phase sync and the chimera state are not commonly observed in swarmalator systems and to the best of our knowledge, have not been studied rigorously (the anti-phase state has been reported previously in~\cite{sar2022swarmalators,ceron2023diverse} and chimera like states were observed in~\cite{hong2021coupling}). The novelty of our work lies in the fact that we have found an anti-phase sync state with all the intra and inter-phase coupling strengths being negative. It can be inferred that the imposed community structure is responsible for this. The chimera state encountered is also due to the interplay between swarmalators belonging to different communities. Although we were not able to provide any mathematical formulation for the chimera state, our model still can be used as a testbed for future works on chimera states in swarmalator systems.

\par We have also conspicuously illustrated the phase transitions by varying the inter-community phase coupling strength $K_3$. The emerging states are characterized in terms of some order parameters. Anti-phase sync state is perceived for a sufficiently small (negative) value of $K_3$ and the sync state is detected for a positive large value of it. We study these two states in detail and provide semi-analytical conditions for achieving these states. We also study the different routes from the anti-phase sync state to the sync state by assuming that the two communities are identical to start with. Moreover, we have established our results when the parameters $J_3$ and $K_3$ are varied simultaneously.

\par We can highlight the limitation of our work by pointing at the inability to explicitly incorporate the spatial dynamics in the analysis of the anti-phase sync and static sync state. This might be wiped out if some simpler type of spatial interaction functions is used other than the power laws used in our model. It also remains to be seen what happens when more than two communities are considered. The model can be simplified by reducing the spatial dimension placing the swarmalators on a ring and then imposing the community interactions. 
 {\color{black} Future works can also be carried out with our model by considering nonidentical swarmalators by drawing frequencies from Gaussian or Lorentzian distributions. Through preliminary inspection, we observed that some of the emerging states that we reported here (static async, static sync) will have their analogous counterparts for nonidentical swarmalators. But for the existence of other states like anti-phase, chimera, etc., a deep and systematic investigation is required.}

\color{black}
\section*{Data availability statement}
The data that support the findings of this study are openly available in the GitHub repository \cite{web_5}.
\color{black}

%%%%%%%%%%%%%%%%%%%%%%%%%%%%%%%%%%%%%%%%%%%%%%%%%%%%%%%

%%%%%%APPENDIX%%%%%%%%
\appendix
\section{Unequal community sizes}
\label{appendixa}
\begin{figure}[htp]
    \centering
    \includegraphics[width=\columnwidth]{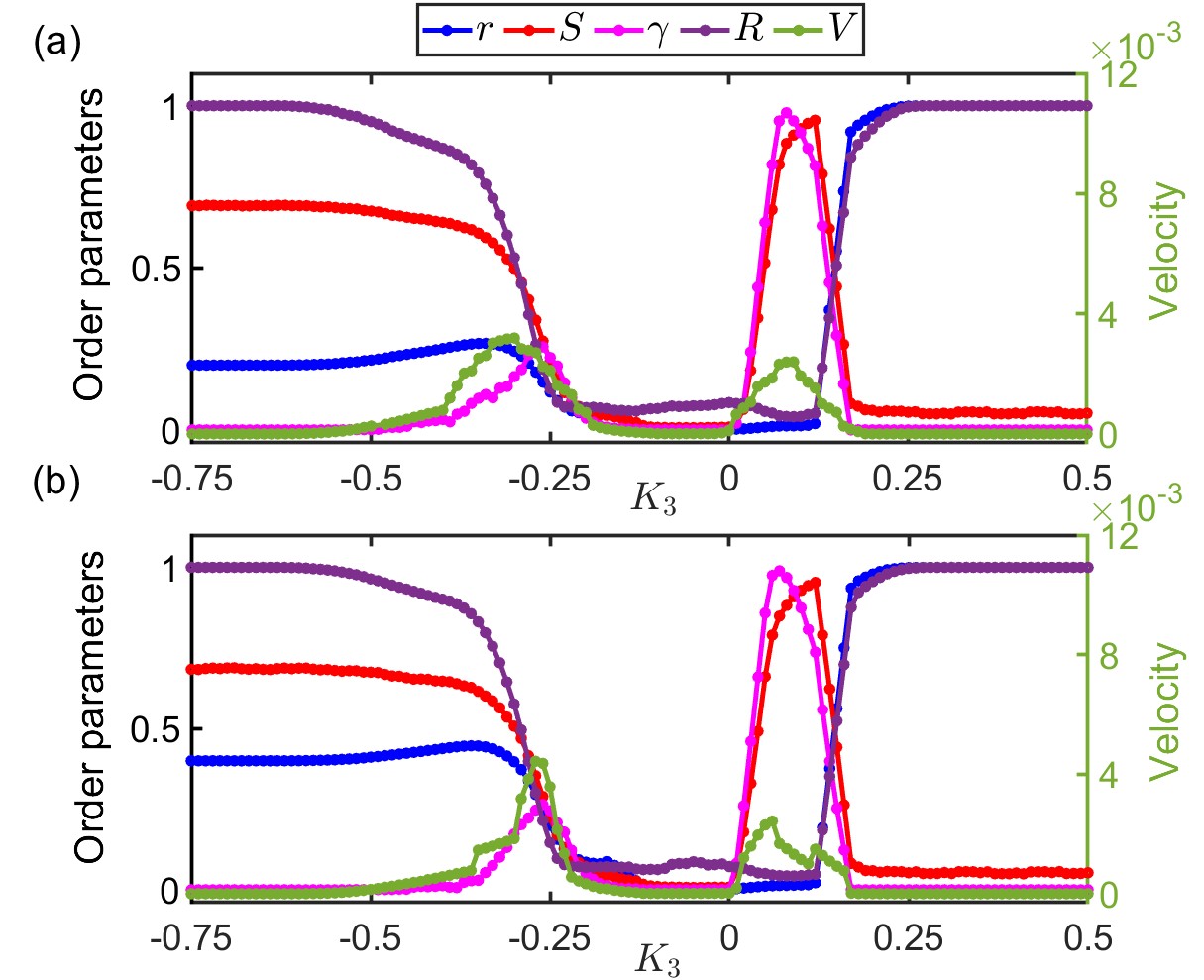}
    \caption{{\bf Phase transition with unequal community sizes.} Simulations are performed for a total population of $N=200$ swarmalators. (a) $p_1=0.6$ and (b) $p_1=0.7$. Other parameter values used are $J_1=J_2=J_3=0.1$ and $K_1=-0.1, K_2=-0.2$. We observe the same states as in Fig.~\ref{op_ppt} with equal community sizes. It establishes the fact that our reported results are robust and independent of the initial distribution of swarmalators in the communities.}
    \label{unequal size}
\end{figure}

In the main text of our paper, we discussed the case where the communities are of equal size and studied different states. Here, we cover the scenario where the two communities have unequal sizes. The total population size is $N$. These swarmalators are distributed in two communities. Let $p_1$ and $p_2$ denote the probabilities that the $i$-th swarmalator belongs to the first, and second communities, respectively. Clearly, $p_1+p_2=1$. For equal community sizes, $p_1$ is essentially equal to $p_2$. Here, we take $p_1 \ne p_2$ so that the communities are unequal in size. We study two cases, one where $p_1=0.6$ and the other one $p_1=0.7$. The parameter values are $J_1=J_2=J_3=0.1$, $K_1=-0.1$, and $K_3=-0.2$, the same values which were used in Sec.~\ref{emergingstates}. In both cases, what we observe that the same qualitative behavior of all the order parameters. As a result, the emerging states remain unaltered. In Fig.~\ref{unequal size}, we have shown the phase transition. In the case of equal population sizes, the order parameter $r$ is approximately zero in anti-phase sync. Due to an equal number of swarmalators in each group, the terms within the summation in Eq.~\eqref{r} nullify each other. But if we choose unequal sub-populations, $r$ has a non-zero value depending on the ratio of swarmalators.

\section{$J_1 \ne J_2$}
\label{appendixb}
\begin{figure}[t]
    \centering
    \includegraphics[width=\columnwidth]{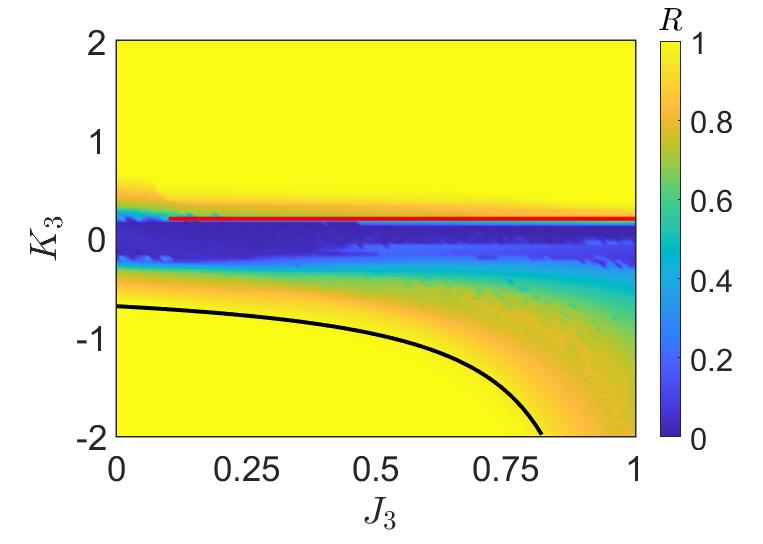}
    \caption{{\bf $J_3$-$K_3$ parameter space for $J_1 \ne J_2$.} Here $K_1=-0.1$, $K_2=-0.2$, $J_1=0.1$, $J_2=0.5$. The system is integrated with $N=200$ swarmalators using Heun's method with step-size $dt=0.01$ for $T=5000$ time units. Order parameter $R$ is calculated with the last $10\%$ data after discarding the transients. Colorbar stands for the value of $R$. Red and black curves are drawn using Eqs.~\eqref{cond1} and~\eqref{cond2}, respectively.}
    \label{j1-ne-j2}
\end{figure}

We study the case where the $J$'s (phase-dependent spatial coupling strengths among communities) are not equal i.e., $J_1 \ne J_2$. For instance, we take $J_1=0.1, J_2 = 0.5$. $K_1$ and $K_2$ are kept fixed at $-0.1$ and $-0.2$, respectively. The resulting behavior is demonstrated through Fig.~\ref{j1-ne-j2}. The overall collective states remain the same. It can be observed if we compare Fig.~\ref{j1-ne-j2} with Fig.~\ref{ps} (where $J_1 = J_2$).

\section{Nonidentical swarmalators}
\label{appendixc}
\begin{figure*}
    \centering
    \includegraphics[width=2\columnwidth]{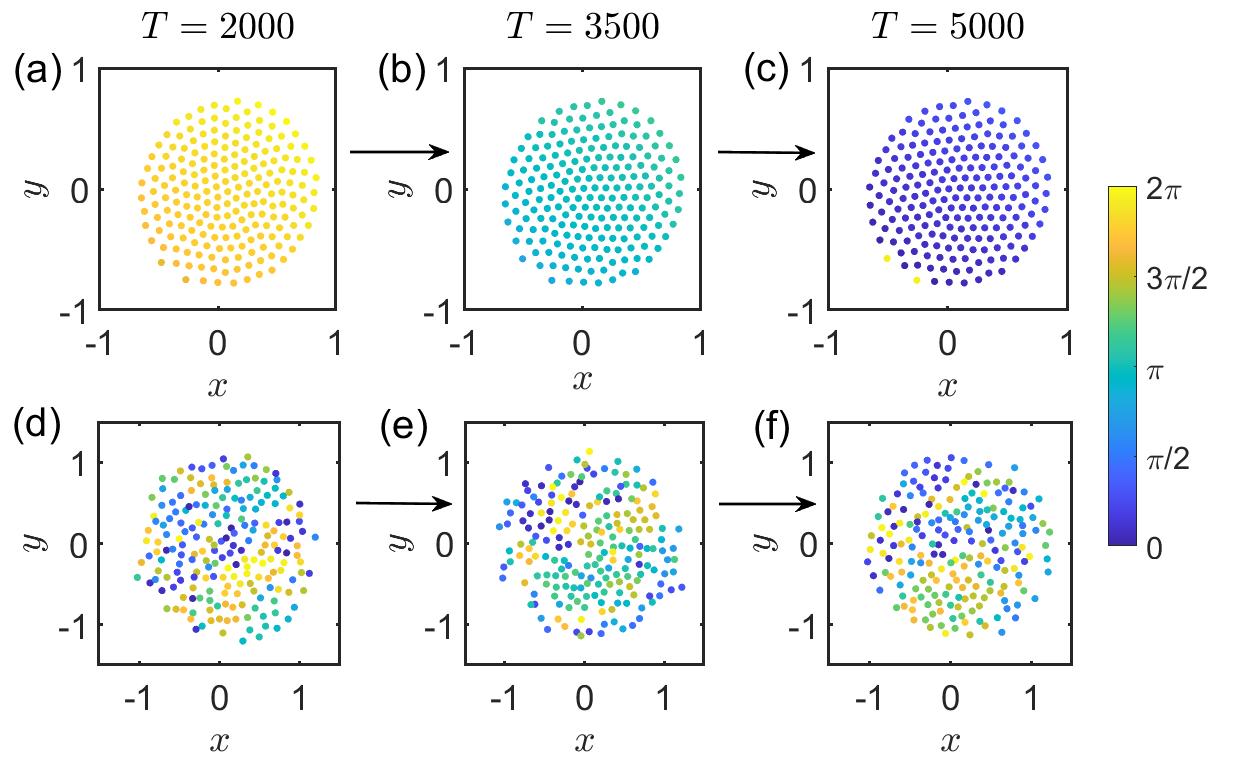}
    \caption{{\color{black}{\bf Sync and async states for nonidentical swarmalators.} We choose $\omega_i$ randomly from Gaussian distribution centered at zero with standard deviation $1$. Simulation parameters: $J_1=J_2=J_3=0.5$. $K_1=-0.1,K_2=-0.2$. $(dt,N)=(0.01,200)$. Snapshots of the sync state are shown at (a) $T=2000$, (b) $T=3500$, and (c) $T=5000$ time units where $K_3=4.0$. The async state is shown for $K_3=0.5$ at (d) $T=2000$, (e) $T=3500$, and (f) $T=5000$ time units. Swarmalators are colored according to their phases. The numbers of swarmalators in both communities are same here.}}
    \label{nonidentical}
\end{figure*}

\color{black}
For our study, we have considered swarmalators with identical frequencies in both the communities, i.e., $\omega_i=\omega$ for $i=1,2,\ldots, N$, and it is further set to zero by moving to a proper reference frame. Here, we draw the frequencies from the Gaussian distribution with zero mean and unit standard deviation to make the nonidentical swarmalators. We observe that the sync state takes place for a larger inter-community phase coupling strength $K_3$ compared to identical swarmalators. The phases never become static and keep evolving with time which is seen via Fig.~\ref{nonidentical}(a)-(c) where snapshots are taken at different time units. For small coupling strength $K_3$, the async state is realized. Here also, the phases are non-stationary. In Fig.~\ref{nonidentical}(d)-(f), snapshots of the async state are shown at $T=2000,3500$, and $5000$ time units, respectively. However, we were unable to detect the emergence of anti-phase and chimera states. A rigorous study through minute exploration of the parameters is needed when one considers nonidentical swarmalators. 
\color{black}
    
\bibliographystyle{apsrev}
\bibliography{reference}

\end{document}